# FROM THERMOSTATICS – TO THE THERMOKINETICS

Etkin V.A.

Present-day thermodynamics has long outgrown the initial frames of the heat-engine theory and transmuted into a rather general macroscopic method for studying kinetics of various transfer processes in their inseparable connection with the thermal form of motion. However its primary notions and mathematical instrument as before based on concepts of thermostatics, to which time, speed and productivity of processes are alien, and on the equations transitory in case of irreversible processes in inequalities. It is offered essentially other approach at which the thermostatics equations follow from thermokinetics of spatially non-uniform systems.

**Introduction.** One of the most attractive features of the thermodynamic method has always been the possibility to obtain a great number of consequences of various phenomena as based on few primary principles. Therefore, it is not by pure accident that all the greatest physicists and many mathematicians of the last century (H.Lorenz, A.Poincaré, M.Planck, W.Nernst, K.Caratheodory, A.Sommerfeld, A.Einstein, M.Born, E.Fermi, J.Neiman, L.Landau, Y.Zeldovich, R.Feynman, etc) in their investigations placed high emphasis on thermodynamics and, based on it, have obtained many significant results. However, thermodynamics have presently lost its peculiar position among other scientific disciplines. In our opinion, one of the reasons of such situation is that thermodynamics has lost its phenomenological nature with considerations of statistical-mechanical character gaining influence in its conceptual basis. As a result, the existing theory of irreversible processes does not reach the rigor and completeness intrinsic for the classic thermodynamic method. Striving for excluding postulates from the grounds of theory dictates the necessity to base thermokinetics on only those statements that are beyond any doubt and construed as axioms [1].

**1. Substantiation of Total energy Conservation Law.** Classic thermodynamics is known to be based on the principle of heat $Q$ and work $W$ equivalence. R. Clausius, the founder of classic thermodynamics, formulated this principle as follows, "In all cases, when heat becomes work in a cyclic process, the amount of the heat expended is proportional to the work done and vice versa, work done is converted into an equivalent amount of heat" [2]. If heat and work are measured in the same units of the international system of units, SI, the equivalence principle may be written as a simple relationship:

$$W_c / Q_c = \oint đW / \oint đQ = 1, \qquad (1)$$

где $W_c$, $Q_c$ – work done and heat supplied for cycle; $đW$, $đQ$ – their elementary amounts for particular parts of the cyclic process under consideration.

Taking into consideration the rule of signs accepted in thermodynamics (the work done by a system and the heat supplied to it are positive) equation (1) becomes:

$$\oint (đQ - đW) = 0. \qquad (2)$$

Clausius was the first who noticed that the above result did not depend on the path of the process under consideration. That allowed him to use a known mathematical theorem of curvilinear integrals. It states that if a curvilinear integral of an arbitrary differential form (in our case $đQ - đW$) becomes zero along any closed path within some space of variables, the integrand represents the exact differential of function of these variables $U$:

$$dU = đQ - đW \text{ or } đQ = dU + đW. \qquad (3)$$

R. Clausius did not concretize the space of variables wherein he considered the curvilinear integral (2) since he had not yet found the heat exchange (entropy) coordinate. Therefore he initially called the function $U$ the *total heat of a body* having construed it as the sum of the heat the body received from outside and the heat released as a result of the *disgregation work* (of dissipative character). That caused some confusion in notions since imparted the status of state function to heat and disgregation work. Therefore a rather heated discussion combusted about the Э function. In particular, W. Thomson recommended the term *mechanical energy of a body in particular state* for the $U$ value. From that time on this function has been referred to as the *internal energy*. Being the state function of a system, that function did not depend on the motion or position of the system relative to the environment. In such a case the isolation of the system from the environment ($Q, W = 0$) left that function invariable. Based on that fact, expression (3) started to be considered in classic thermodynamics as a particular case of the *energy conservation law* called the first law of thermodynamics.

Since classic thermodynamics from its origin has always been restricted to describing the behavior of internally equilibrium (spatially homogeneous) systems with parameters the same for all of the system parts, all kinds of work such a system could do had the *unordered* character [2]. Here comes, in particular, the uniform compression work $dW_p$ expressed through the product of the absolute pressure $p$ and the volume variation $dV$. In such a case the Э function could actually be construed as the dissipated part of energy, which corresponded to the law of energy conservation in the form of (4). However, in the more general case of non-equilibrium and especially spatially heterogeneous systems this is far from being so. In particular, the chemical and nuclear energy of homogeneous systems is also partly convertible into other forms despite they do not either depend on the position of the system relative to surrounding bodies, i.e. relate to internal energy. The situation became even more complicated with changing to the spatially heterogeneous systems to be studied, in particular, to the so-called *extended* systems with the environment included. Such systems can do some useful work before internal equilibrium has set in there. These systems may comprise also polarized and magnetized bodies located in external force fields. In all those cases the internal energy $U$ ceased being that "dissipated" part of energy meant in (3).

All this impels to search for a more general substantiation of the law of its conservation. To this end let us consider the results of those experiments on definition of the heat and work equivalence principle, which related to non-equivalent systems with relaxation processes running therein. Their specific character was such that heat was obtained there by friction (dissipation). These include classic experiments by Joule, in particular, the experiment with calorimeter and agitator driven by dropping weight; also his experiment with the Proney brake that brakes the drum calorimeter (1843-1878); Girn's experiments with lead flattening on anvil with drop hammer (1859); Lenz's experiments with solenoid discharging to active resistance in vessel calorimeter (1972) and many other experiments involving battery charging, gas transfer between vessels, electrolyte decomposition, etc [3]. Those experiments had such a result that a system disturbed from equilibrium by a mechanical (ordered) work $W_i^e = W_i^e(\mathbf{r}_i)$ done on it returned to the initial equilibrium state after a heat amount $Q$ strictly equivalent to the work had been removed from the system. Taking the work of both ordered and unordered character (according to the above classification) into consideration means the necessity to extend the space of variables wherein the above considered cyclic processes take place. It is easy to reveal in this case that integrand (2) constitutes a state function in the space of variables ($\Theta_i, \mathbf{r}_i$), i.e. a more general one than the system energy. This function depends on both the internal $\Theta_i$ and external $\mathbf{r}_i$ system coordinates, i.e. constitutes the sum of the external and internal system energies. Such energy is usually called the *total* energy of the system. The decrease of the function $Э = Э(\Theta_i, \mathbf{r}_i)$ defines the sum of all (ordered and unordered) works the system do:

$$-dЭ(\Theta_i, \mathbf{r}_i) = \Sigma_i\, dW_i^a + \Sigma_i\, dW_i^e. \tag{4}$$

According to this expression the energy of a system, in the absence of external impacts on the system ($\Sigma_i dW_i^a = 0$, $\Sigma_i dW_i^e = 0$), remains invariable at any variations of its state. In other words, the *energy of an isolated system is constant*. Thus the generalization of the heat and work equivalence principle to non-equilibrium systems directly leads to the law of conservation of "total" energy as a state function for the entire set of interacting (mutually moving) bodies. However, for such a system (isolated) all its energy is *internal*. This fact reveals the imperfection of dividing the energy into *external* and *internal*. From the position of energodynamics considering the entire set of interacting bodies as a single non-equilibrium whole it is more important that its energy be measured in an *own* (absolute) reference frame not connected with the state of any of the bodies within the environment[1]. Since the term *system energy* with regard to the function $Э(Θ_i,\mathbf{r}_i)$ unambiguously tells the energy belongs to the system itself, the terms "total", "external", "internal", etc energies become superfluous. This allows focusing on other properties of energy and its other components which characterize its conversion capacity.

## 2. Extension of Variables Space with Introduction of Spatial Heterogeneity Parameters for System as a Whole

The fact that relaxation vector processes (temperature, pressure, concentration, etc equalization) run in non-equilibrium systems requires introducing specific parameters of spatial heterogeneity characterizing the state of continuums as a whole. To do so, it is necessary, however, to find a way how to change over from the density (fields) distribution functions $ρ_i = dΘ_i/dV$ of some extensive physical values $Θ_i$ to the parameters of the system as a whole, which thermodynamics operates with. This change may be conducted in the same way as used in mechanics to change over from motion of separate points to system center-of-mass motion. To better understand such a change, let us consider an arbitrary continuum featuring non-uniform density distribution $ρ_i = ρ_i(\mathbf{r},t)$ of energy carriers[2] over the system volume $V$. Fig.1 illustrates the arbitrary density distribution $ρ_i(\mathbf{r},t)$ as a function of spatial coordinates (the radius vector of a field point $\mathbf{r}$) and time $t$. As may be seen from the figure, when the distribution $Θ_i$ deviates from that uniform (horizontal line), some amount of this value (asterisked) migrates from one part of the system to other, which displaces the center of this value from the initial $\mathbf{R}_{i0}$ to a current position $\mathbf{R}_i$.

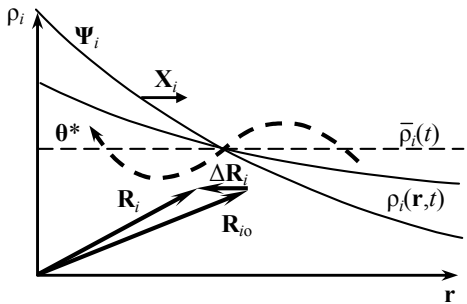

Fig.1. To Generation of Distribution Moment

Position of the center of a particular extensive value $Θ_i$ defined by the radius vector $\mathbf{R}_i$ is given by a known expression:

$$\mathbf{R}_i = Θ_i^{-1} \int ρ_i(\mathbf{r},t)\,\mathbf{r}dV, \quad (i = 1,2,…,n) \qquad (5)$$

For the same system, but in a homogeneous state, the $Θ_i$ center position $\mathbf{r}_{i0}$ may be derived if factoring $ρ_i = ρ_{i0}(t)$ in equation (1) outside the integral

$$\mathbf{R}_{i0} = Θ_i^{-1}\int_V ρ_{i0}(t)\mathbf{r}dV = V^{-1}\int_V \mathbf{r}dV. \qquad (6)$$

---

[1] Otherwise, should the Energy conservation be violated, the Energy of a system would vary with the state variation of these bodies despite the absence of Energy exchange with the system.

[2] The Energy carrier is construed as a material carrier of the $i^{th}$ Energy component, which quantitative measure is the physical value $Θ_i$. So the mass $M_k$ of the $k^{th}$ substance is a carrier of the rest Energy; the charge $Θ_e$ – a carrier of the electrostatic Energy of the system; the component momentum $M_k v_k$ – a carrier of its kinetic Energy, etc.

Thus the state of a heterogeneous system features the emergence of specific "distribution moments" $\mathbf{Z}_i$ of the energy carriers $\Theta_i$:

$$\mathbf{Z}_i = \Theta_i(\mathbf{R}_i - \mathbf{R}_{i0}) = \int_V [\rho_i(\mathbf{r},t) - \rho_{i0}(t)]\mathbf{r}dV. \qquad (7)$$

The electrical displacement vector $\mathbf{D} = \Theta_e \Delta \mathbf{R}_e$ is one of such moments with $\Theta_e$ as electrical charge and $\Delta \mathbf{R}_e$ as displacement of its center.

Expression (7) most evidently manifests that the parameters $\mathbf{Z}_i$ of spatial heterogeneity are additive values and summed up providing the $\rho_{i0}(t)$ value remains the same in various parts of a heterogeneous system. This follows from the conservation of integral (7) at its partition into parts with a volume $V' < V$ [1]. However, these parameters become zero at "contraction" of the system to a material point, when $\rho_i(\mathbf{r},t) \to \rho_{i0}(t)$. This stands in absolute conformity with the degrees-of-freedom theorem because the processes of density redistribution $\rho_i(\mathbf{r},t)$ are absent in material points. And once again this confirms the fact that an entity of continuum elements considered as a system, non-equilibrium in whole, possesses additional degrees of freedom.

For any part of a homogeneous isolated system the $\mathbf{r}_{i0}$ value remains unvaried since running of any processes is herein impossible. Therefore the $\mathbf{r}_{i0}$ may be accepted for such systems as a reference point $\mathbf{r}$ or $\mathbf{r}_i$ and set equal zero ($\mathbf{r}_{i0} = 0$). In this case the vector $\mathbf{r}_i$ will define a displacement of the $\Theta_i$ center from its position for the system being in internal equilibrium state, and the moment of distribution of a particular value $\Theta_i$ in it will become:

$$\mathbf{Z}_i = \Theta_i \mathbf{R}_i \qquad (8)$$

Herein the moment $\mathbf{Z}_i$ becomes an absolute extensive measure of the system heterogeneity with respect to one of the system properties – like such absolute parameters of classic thermodynamics as mass, volume, entropy, etc.

Explicitly taking into account the spatial heterogeneity of systems under investigation is decisive in further generalization of the thermodynamic investigation method to non-equilibrium systems. As a matter of fact, this is the spatial heterogeneity (heterogeneity of properties) of natural objects that causes various processes running in them. This implies the exclusive role the distribution moments $\mathbf{Z}_i$ play as a measure for deviation of a system in whole from internal equilibrium of the $i^{th}$ kind. Introducing such parameters allows precluding the major drawback of non-equilibrium thermodynamics, viz. lack of extensive variables relating to the gradients of temperature, pressure, etc. Classic thermodynamics is known to have crystallized into an independent discipline after R. Clausius succeeded in finding a coordinate (entropy) related to temperature in the same way as pressure to volume and thus determinately described the simplest thermo-mechanical systems. The distribution moments $\mathbf{Z}_i$ play the same part in thermokinetics coming into being. As will be shown later, these relate to the main parameters introduced by non-equilibrium thermodynamics – thermodynamic forces, in the same way as the generalized potentials to the coordinates in equilibrium thermodynamics. These are the distribution moments which make the description of heterogeneous media a deterministic one thus enabling introducing in natural way the concept of generalized velocity of some process (flow) as their time derivatives. They visualize such parameters as the electrical displacement vectors in electrodynamics and generalize them to phenomena of other physical nature. In mechanics the $\mathbf{Z}_i$ parameters have the dimension of action ($\Theta_i$ – momentum of a body, $\mathbf{R}_i$ – its displacement from equilibrium position), imparting physical meaning to this notion. These are the parameters which allow giving the analytic expression to the system working capacity having thus defined the notion of system energy. Using such parameters provides a clear view of the degree of system energy order, enables proposing a universal criterion of the non-equilibrium system evolution, etc. Paraphrasing a M. Planck's statement regarding entropy one may positively say that the distribution moments are exactly the parameters entire non-equilibrium thermodynamics is "standing and falling" with.

---

[1] With symmetrical density $\rho_i(\mathbf{r},t)$ distributions for whatever parameter, e.g., fluid-velocity profiles in tubes, expression (7) should be integrated with respect to annular, spherical, etc layers with $V' > 0$, wherein the function $\rho_i(\mathbf{r},t)$ is monotone increasing or decreasing.

### 3. Coordinates of Non-Equilibrium Redistribution and Reorientation Processes

The moments of distribution (7) contain vectors of displacement $\mathbf{R}_i$, each of which can be expressed product of a basic (individual) vector $\mathbf{e}_i$, characterising its direction, on module $R_i = |\mathbf{R}_i|$ this vector. Therefore the complete variation of the displacement vector $\mathbf{r}_i$ may be expressed as the sum of two summands:

$$d\mathbf{R}_i = \mathbf{e}_i dR_{ii} + R_i d\mathbf{e}_i, \qquad (9)$$

where the augend $\mathbf{e}_i dr_i = d\mathbf{r}_i$ characterizes elongation of the vector $\mathbf{R}_i$, while the addend $R_i d\mathbf{e}_i$ – its turn.

Let us express now the $d\mathbf{e}_i$ value characterizing the variation of the distribution moment direction in terms of an angular displacement vector $\boldsymbol{\varphi}$ normal to the plane of rotation formed by the vectors $\mathbf{e}_i$ and $d\mathbf{e}_i$. Then the $d\mathbf{e}_i$ will be defined by the external product $d\boldsymbol{\varphi}_i \times \mathbf{e}_i$ of vectors $d\boldsymbol{\varphi}_i$ and $\mathbf{e}_i$, so the addend in (1.6.1) will be $\Theta_i R_i d\mathbf{e}_i = d\boldsymbol{\varphi}_i \times \mathbf{Z}_i$. Hence, expression of full differential of the distribution moments looks like:

$$d\mathbf{Z}_i = (\partial \mathbf{Z}_i / \partial \Theta_i) d\Theta_i + (\partial \mathbf{Z}_i / \partial \mathbf{r}_i) d\mathbf{r}_i + (\partial \mathbf{Z}_i / \partial \boldsymbol{\varphi}_i) d\boldsymbol{\varphi}_i. \qquad (10)$$

According to the degrees-of-freedom theorem this means that any state function describing a heterogeneous system in whole are generally defined by also the full set of variables $\Theta_i$, $\mathbf{r}_i$ and $\boldsymbol{\varphi}_i$. Since further resolution of the vector $\mathbf{Z}_i$ is impossible, expression (7) indicates there are three categories of processes running in heterogeneous media, each having its own group of independent variables. The first-category processes running at $\mathbf{R}_i = $ const involve the uniform variation of the physical value $\Theta_i$ in all parts of the system. Such processes resemble the uniform rainfall onto an irregular (in the general case) surface. Here comes, in particular, the pressure field altered in liquid column with variation of free-surface pressure. These processes also cover phase transitions in emulsions, homogeneous chemical reactions, nuclear transformations and the similar scalar processes providing the composition variations they induce are the same in all parts of the system. We will call them hereinafter the *uniform processes* regardless of what causes the increase or decrease in amount of whatever energy carrier $\Theta_i$ (and the momentum associated) – either the external energy exchange or internal relaxation phenomena. These processes comprise, as a particular case, the reversible (equilibrium) processes of heat exchange, mass exchange, cubic strain, etc, which, due to their quasi-static nature, practically do not disturb the system spatial homogeneity.

Processes described by the addend in (7) run with the $\Theta_i$ parameters being constant and consist in their redistribution among the parts (zones) of a heterogeneous system. They involve decreasing, e.g., the entropy $S'$, mass $M'$, its momentum $P'$, its volume $V'$, etc, in some parts of the system and by increasing the same in other parts. Such processes are associated with the $\Theta_i$ value center position variation $\mathbf{R}_i$ within the system and resemble the migration of fluids from one part of a vessel into another. Therefore we will call them the *redistribution* processes. Such processes are always non-equilibrium even if they run infinitely slowly (quasi-statically) since the system remains spatially heterogeneous in this case. State modifications of such a kind are caused by, e.g., the useful external work of external forces, the non-equilibrium energy exchange processes that induce non-uniform variation of the $\Theta_i$ coordinates inside the system, and the vector relaxation processes involving equalization of temperature, pressure, chemical and other system potentials. All processes of such a kind feature a directional (ordered) character, which distinguishes the useful work from the work of uniform (quasi-static) introduction of substance, charge, etc, or the expansion work. According to (1.3.2) the coordinates of the processes pertaining to this category are understood as the displacement vectors $\mathbf{R}_i$. These coordinates should be attributed to the *external parameters* of the system since they characterize the *position* of the energy carrier $\Theta_i$ center in whole relative to external bodies (the environment) just as the center of mass $\mathbf{R}_m$ of the system or its center of inertia $\mathbf{R}_w$.

There are also the processes of *reorientation* of magnetic domains, electrical and magnetic dipoles, axes of rotation of bodies, etc., running in a number of systems, e.g., in ferromagnetic materials. The micro-world manifests them in, e.g., the unified spin-orientation arrangement' the macro-systems – in the spontaneous magnetization of ferromagnetic materials, while the mega-world – in the close-to-equatorial plane alignment of the galaxies' spirals, asteroidal belts, orbits of the primary planets and their satellites, etc. The systems with processes of such a kind will hereafter be called, for short, *oriented*. These include also the bodies with shape anisotropy. The reorientation processes are not reducible to the transfer and redistribution processes either. This means that the coordinate of such kind a process is a variation of the angle $\varphi_i$ characterizing the orientation of distribution moment $\mathbf{Z}_i$ of the system as a whole.

Thus, all processes running in heterogeneous systems may be broken down into three groups: *uniform, redistribution* and *reorientation processes*, which coordinates are, respectively, variables $\Theta_i$, $\mathbf{r}_i$ and $\varphi_i$. This fundamentally distinguishes thermokinetics from classic thermodynamics and the theory of irreversible processes, where the state of a system is defined by exclusively a set of thermostatic variables $\Theta_i$.

The undertaken expansion of the space of variables by introducing the vectors of displacement $\mathbf{R}_i$ makes it possible to cover not only quantitative, but as well *qualitative* variations of energy in various forms. The fact that *vector processes* run in systems along with *scalar processes* means that both the *ordered* $W^e$ and *unordered* $W^{un}$ works are generally done in such systems. It becomes clear that the irreversibility of real processes associated with the energy dissipation (i.e. with losing the capacity for ordered work) becomes apparent in the process *scalarization*, i.e. in losing vector character of the process. Furthermore, a possibility appears to further distinguish between the *energy transfer* processes (i.e. the energy transfer between bodies in the same form) and the *energy conversion* processes (i.e. the energy conversion from one form into another)[1].

### 4. Introduction into basic equation of energodynamics force and its moment

Let us consider the consequences ensuing from the fact itself of existing the system energy $Э(\Theta_i, \mathbf{r}_i)$ as a function of the quite certain set of arguments (state coordinates) As shown above, the energy of a heterogeneous system as a function of its state is generally expressed as $Э = Э[\mathbf{Z}_i(\Theta_i, \mathbf{R}_i, \varphi_i)]$, where $i = 1, 2, …, n$ – number of energy components equal to the maximal number of independent processes for some of their categories (uniform processes, redistribution and reorientation processes). This means that the exact differential of energy may be expressed by the following relationship [4]:

$$dЭ = \Sigma_i(\partial Э/\partial\Theta_i)d\Theta_i + \Sigma_i(\partial Э/\partial\mathbf{r}_i)d\mathbf{r}_i + \Sigma_i(\partial Э/\partial\varphi_i)d\varphi_i. \qquad (11)$$

Derivatives of some system parameters ($Э$) with respect to other ones ($\Theta_i$, $\mathbf{r}_i$, $\varphi_i$) are also system parameters. Therefore denoting them as:

$$\Psi_i \equiv (\partial Э/\partial\Theta_i); \qquad (12)$$

---

[1] As will be shown hereinafter, the Energy transfer is associated with unordered work done, whereas the Energy conversion – with ordered work.

$$\mathbf{F}_i \equiv -(\partial Э/\partial \mathbf{r}_i); \tag{13}$$

$$\mathbf{M}_i \equiv -(\partial Э/\partial \boldsymbol{\varphi}_i), \tag{14}$$

gives the fundamental identity of energodynamics in the form:

$$dЭ \equiv \Sigma_i \Psi_i dΘ_i - \Sigma_i \mathbf{F}_i \cdot d\mathbf{r}_i - \Sigma_i \mathbf{M}_i \cdot d\boldsymbol{\varphi}_i, \tag{15}$$

For isolated systems the right-hand member of identity (9) becomes zero. For systems not changing its spatial orientation ($\boldsymbol{\varphi}_i$ = const) the two last terms in (9) may be combined, then the fundamental identity of energodynamics becomes:

$$dЭ \equiv \Sigma_i \Psi_i dΘ_i - \Sigma_i \mathbf{F}_i \cdot d\mathbf{r}_i. \tag{16}$$

Identities (15) and (16) are nothing else but a result of the joint definition of the related parameters $\Psi_i$ and $Θ_i$, $\mathbf{F}_i$ and $\mathbf{r}_i$ or $\mathbf{Z}_i$, $\mathbf{M}_i$ and $\boldsymbol{\varphi}_i$. To clarify the physical meaning of the parameters thus introduced, let us first consider the particular case of internally equilibrium (spatially homogeneous) and stationary thermo-mechanical systems. Such simplest systems may be instantiated as the working media of heat engines in the vaporous or gaseous state. They have two degrees of freedom – thermal and mechanical, i.e. the capacity for the heat exchange $Q$ and the uniform expansion work $W_{ex}$. Due to the absence of redistribution and reorientation processes in homogeneous systems ($d\mathbf{r}_i$, $d\boldsymbol{\varphi}_i = 0$) the parameters $\Psi_i$ are the same for all points of such a system and equal to their local values $\psi_i$, so that identity (9) goes over into a joint equation of the first and second laws of thermodynamics for closed systems:

$$dU = \Sigma_i \Psi_i dΘ_i = TdS - pdV. \tag{17}$$

Since the variation of the coordinates $Θ_i$ in an equilibrium system is caused by exclusively the external heat exchange (their internal sources are absent), the terms of this relationship characterize, respectively, the elementary heat exchange in the system $đQ = TdS$ and the elementary expansion work $đW_{ex} = pdV$. In this case the parameters $\Psi_i$ acquire the meaning of the absolute temperature $T$ and absolute pressure $p$. In the more general case of spatially heterogeneous systems the parameters $\Psi_i$ are, as will be shown hereinafter, the generalized local potentials $\psi_i$ averaged by mass in all elements of the system.

To clarify the meaning of the terms of the second sum in (9), we must take into account that they correspond to the redistribution processes running at constant parameters $Θ_i$ and $\boldsymbol{\varphi}_i$, i.e. with invariable direction of the unit vector $\mathbf{r}_i$. Then $(\partial Э/\partial \mathbf{Z}_i) = Θ_i^{-1}(\partial Э/\partial \mathbf{r}_i) = -\mathbf{F}_i/Θ_i = -\mathbf{X}_i$, then the $\mathbf{X}_i$ thermodynamic forces thus introduced are actually the specific forces in their usual (Newtonian) meaning, i.e. the forces $\mathbf{F}_i$ per unit of the value $Θ_i$ they transfer. These are, in particular, the specific mass, bulk and surface forces, for which the $Θ_i$ value is construed as, respectively, mass $M$, volume $V$ and surface $f$ of the body. This category also includes the Lorenz force $\mathbf{F}_e$ related to the electric charge $Θ_e$ transferred. Using them enables representation of work by two equivalent expressions:

$$đW_i^e = \mathbf{F}_i \cdot d\mathbf{r}_i = \mathbf{X}_i \cdot d\mathbf{Z}_i. \tag{18}$$

The work described by expression (18) may be mechanical, thermal, electrical, chemical, etc (depending on nature of the forces to overcome); external or internal (depending on where the forces arise – either in the system itself or outside); useful or dissipative (depending on what the work involves – either purposeful conversion of energy or its dissipation).

Lastly, the terms of the third sum in (15) correspond to the reorientation processes running with constant $Θ_i$ and $\mathbf{r}_i$. In this case $\mathbf{F}_i \cdot d\mathbf{r}_i = \mathbf{F}_i \cdot [d\boldsymbol{\varphi}_i, \mathbf{r}_i]$, and the parameter $\mathbf{M}_i$ acquires the meaning of a torque from the force $\mathbf{F}_i$:

$$\mathbf{M}_i = \mathbf{F}_i \times \mathbf{R}_i \tag{19}$$

This "torsion" torque is advisable to be called the "orientation" torque in the case it becomes zero when the direction of the force $\mathbf{F}_i$ coincides with the direction of the displacement vector $\mathbf{R}_i$.

The fundamental identity of energodynamics thus obtained is valid regardless of what causes the variation of the parameters $\Theta_i$, $\mathbf{r}_i$ and $\mathbf{\varphi}_i$ – either the external heat exchange or the internal (including relaxation) processes. Therefore it is applicable to *any processes* (both reversible and irreversible). At the same time it is most detailed of all the relationships connecting the parameters of spatially heterogeneous systems since it allows for *any possible categories of processes* running in such systems.

Let us pay attemtion now to the fundamental difference between ordered and unordered works described in this expression by the variables of scalar and vector character. For this let us consider first some heterogeneous system consisting of two subsystems with the parameters $\Psi_i{'}$, $\Theta_i{'}$ and $\Psi_i{''}$, $\Theta_i{''}$. If such a system is homogeneous as a whole ($\mathbf{X}_i, \mathbf{M}_i = 0$) and isolated ($d\Э = 0$), expression (15) for it takes the form:

$$d\Э = \Psi_i{'} d\Theta_i{'} + \Psi_i{''} d\Theta_i{''} = 0. \tag{20}$$

Hence it follows that in the process of redistribution of the energy carrier $\Theta_i$ between the parts of such a system $d\Theta_i{'} = - d\Theta_i{''}$ the value of the $i^{th}$ energy form $U_i$ therein remains invariable, i.e. only a transfer of energy occurs in this form across the border between these parts. We called such an energy exchange running without energy form variation as the *energy transfer* for short.

Another kind are the processes described by the terms of $\mathbf{X}_i \cdot d\mathbf{Z}_i$ type or $\mathbf{M}_i \cdot d\mathbf{\varphi}_i$ as their variety. If a system is heterogeneous, i.e. $\mathbf{X}_i = -\nabla\Psi_i \neq 0$ and $d\mathbf{Z}_i = \Theta_i d\mathbf{r}_i \neq 0$, then

$$\mathbf{X}_i \cdot d\mathbf{Z}_i = - \Theta_i (d\mathbf{r}_i, \nabla)\Psi_i = - \Theta_i d\Psi_i(\mathbf{r}_i), \tag{21}$$

where $\Psi_i(\mathbf{r}_i)$ is the potential of some part in the heterogeneous system, which varies with part-to-part transfer within the system, i.e. should be considered as a function of system position $\mathbf{r}_i$ in the field of the $\Psi_i$ potential. Thus the terms $\mathbf{X}_i \cdot d\mathbf{Z}_i$ describe the $i^{th}$ energy form variations caused by the above redistribution of the energy carrier $\Theta_i$ if kept in the system as a whole. In accordance with the energy conservation law this is possible only as a result of other energy forms converted into the $i^{th}$ form. Therefore ordered work is always associated with the *energy conversion* process.

### 5. Introduction of the rate and productivity of real processes in equations of thermokinetics

Due to the fact that energodynamics rejects in its grounds the process idealization expressed in such notions as "quasi-static" (infinitely slow), "equilibrium" and "reversible" a possibility appears to introduce time as a logically consistent physical parameter into its equations. For that it is enough to rewrite identity (15) in the form containing total derivatives of the state parameters earlier introduced with respect to time $t$:

$$d\Э/dt \equiv \Sigma_i \Psi_i d\Theta_i/dt - \Sigma_i \mathbf{F}_i \cdot \mathbf{v}_i - \Sigma_i \mathbf{M}_i \cdot \mathbf{\omega}_i. \tag{22}$$

Here $\mathbf{v}_i \equiv d\mathbf{r}_i/dt = \mathbf{e}_i dr_i/dt$ – translation velocity of the energy carrier $\Theta_i$; $\mathbf{\omega}_i \equiv d\mathbf{\varphi}_i/dt$ – angular velocity of its reorientation (or rotation). In the particular case, when the parameter $\Theta_i$ means mass of a system, the values $\mathbf{v}$ and $\mathbf{\omega}$ characterize its linear and angular velocity as a whole. For the future it is quite important to obtain the local statement of this identity true for any element of the continuum. For this purpose let us apply equation (15) first to the system where redistribution processes are absent. Then the second and the third sums in (15) disappear, and the identity becomes:

$$d\Э/dt \equiv \Sigma_i \Psi_i d\Theta_i/dt. \qquad (23)$$

In the systems this equation represents the variation of the $\Theta_i$[1] parameters is caused by exclusively the transfer of some amount of energy carrier across the system borders. This allows representing the behavior of these parameters in the time domain by a known expression:

$$d\Theta_i/dt = -\int \mathbf{j}_i^e \cdot d\mathbf{f}, \qquad (24)$$

where $\mathbf{j}_i^e = \rho_i \mathbf{v}_i$ – local density of flow of the energy carrier $\Theta_i$ through a vector element $d\mathbf{f}$ of the closed surface $f$ in the direction of external normal $\mathbf{n}$; $\mathbf{v}_i$ – velocity of energy carrier transfer through the system surface element $d\mathbf{f}$ in stationary reference frame (Fig. 2).

Substituting (25) into (24) gives:

$$d\Э/dt = -\Sigma_i \Psi_i \int \mathbf{j}_i^e \cdot d\mathbf{f}. \qquad (25)$$

This equation is evidently a particular case of the more general expression

$$d\Э/dt = -\Sigma_i \int \psi_i \mathbf{j}_i^e \cdot d\mathbf{f}, \qquad (26)$$

when the local value $\psi_i$ of generalized potential $\Psi_i$ is the same for all system points and may therefore be factored outside the integral sign. The product $\psi_i \mathbf{j}_i$ is the $i^{th}$ component of the energy flow density $\mathbf{j}_e = \Sigma_i \psi_i \mathbf{j}_i$ through an element $d\mathbf{f}$ of the system surface $f$. Therefore changing in (25) to the integral taken over system volume as based on the Gauss-Ostrogradsky's theorem we come to the expression for the law of energy conservation for an arbitrary continuum area, which was proposed by N. Umov in 1873:

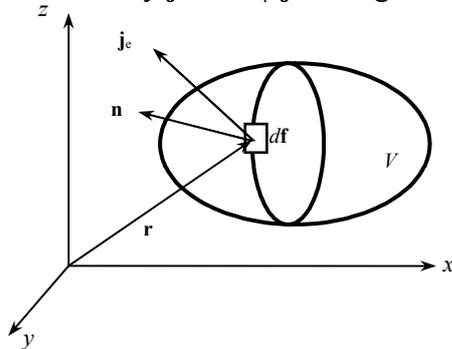

$$d\Э/dt = -\int \nabla \cdot \mathbf{j}_e dV. \qquad (27)$$

Fig. 2. Energy Flow across System Borders

According to this expression the system energy variation equals the amount of energy having passed across the system borders for that particular time. Or according to Umov himself, "energy flow…is caused by energy admission or release a medium provides across its borders". It should be noted that the validity of this statement is by no means restricted to the mechanical energy N. Umov meant.

This equation may be developed by representing the energy flow divergence $\nabla \cdot \mathbf{j}_e = \Sigma_i \nabla \cdot (\psi_i \mathbf{j}_i^e)$ as a sum of two terms $\Sigma_i \psi_i \nabla \cdot \mathbf{j}_i^e + \Sigma_i \mathbf{j}_i^e \cdot \nabla \psi_i$:

$$d\Э/dt = -\Sigma_i \int \psi_i \nabla \cdot \mathbf{j}_i^e dV + \Sigma_i \int \mathbf{x}_i \cdot \mathbf{j}_i^e dV, \qquad (28)$$

where

$$\mathbf{x}_i \equiv -\nabla \psi_i \qquad (29)$$

---

[1] From the physical standpoint the value $\Theta_i$ that is actually the extensive measure of particular kind interaction (Energy) carrier is advisable to be called for short the *Energy carrier*. This will facilitate the understanding of many processes under investigation.

is a local motive force of the $i^{th}$ process expressed as negative gradient of generalized potential and named in the theory of irreversible processes as the "thermodynamic force in its energetic representation" [4].

Equation (29) enables clarifying the meaning of the "global" variables $\Psi_i$ and $\mathbf{X}_i$ introduced earlier for a system in whole. Taking into account that volume or mass elements in continuums do not change their spatial orientation ($d\boldsymbol{\varphi}_i = 0$) may be expressed in the form:

$$d Э/dt \equiv \Sigma_i \Psi_i d\Theta_i/dt - \Sigma_i \mathbf{X}_i \cdot \mathbf{J}_i, \qquad (30)$$

where

$$\mathbf{J}_i \equiv (\partial \mathbf{Z}_i/\partial t)_\varphi = \Theta_i \mathbf{e}_i d\mathbf{r}_i/dt = \Theta_i \mathbf{v}_i, \qquad (31)$$

i.e. are total flows of displacement (transfer) of the $i^{th}$ energy carrier $\Theta_i$.

These flows at $d\boldsymbol{\varphi}_i = 0$ may be expressed in terms of their densities $\mathbf{j}_i \equiv \rho_i d\mathbf{r}_i\,dt$ through the evident relationship:

$$\mathbf{J}_i \equiv \Theta_i d\mathbf{r}_i/dt = \int (d\mathbf{r}_i/dt) d\Theta_i = \int \mathbf{j}_i\, dV. \qquad (32)$$

It is easy to see that the flows $\mathbf{J}_i$ differ in their dimensions from the more usual notion of flow rate and in their meaning as per (31) are closer to the "generalized momentum" $\mathbf{P}_i = \Theta_i \mathbf{v}_i$ of the $i^{th}$ energy carrier $\Theta_i$ for a system in whole. Such flows play an important role in many phenomena. These are, e.g., the vector flows of electric displacement in a system with the volume $V$ defined by the product of the system free charge $\Theta_e$ and the velocity of its center displacement in the free charge redistribution processes. This is the value, to which the following parameters are proportional: magnetic field induction vector (Biot-Savart's law), Thomson–Joule heats in conductors and thermo-elements, electromagnetic force driving a conductor with current (Ampere's law), etc. We will hereinafter be referring to them time and again when dealing with the transfer and conversion of energy in any forms, which will confirm the necessity and usefulness of generalizing the Maxwell's displacement current concept to phenomena of other nature.

To find the relation between the "global" (pertaining to a system in whole) and the local thermodynamic forces $\mathbf{x}_i$, let us take into account that the parameters $\Psi_i$ in identity (15) are defined for the coordinates $\mathbf{Z}_i$ being constant, i.e. for the difference $\rho_i(\mathbf{r},t) - \bar{\rho}_i(t)$ invariable in all points of the system volume $V$. From this it follows that in the expression

$$d\Theta_i/dt = \int (d\theta_i/dt)\, \rho dV, \qquad (33)$$

the specific parameters $\theta_i$ vary uniformly in all parts of the system, so that $d\theta_i/dt$ may be factored outside the integral sign. Hence,

$$\Psi_i = M^{-1} \int \psi_i dM, \qquad (34)$$

being the system mass-averaged value of the local potential $\psi_i$. Similarly proceeding from the invariance of the process power $N_i = \mathbf{X}_i \cdot \mathbf{J}_i$ when representing it in terms of the local and global parameters

$$\mathbf{X}_i = \mathbf{J}_i^{-1} \int \mathbf{x}_i \cdot \mathbf{j}_i\, dV, \qquad (35)$$

gives that the "global" thermodynamic force $\mathbf{X}_i$ is some averaged value of the local thermodynamic force $\mathbf{x}_i \equiv -\nabla\psi_i$.

The relationship thus obtained between the local variables the field theories operate with and the thermodynamic parameters characterizing the state of a continuum in whole opens the possibility of describing their properties from the positions of energodynamics. In this case particular importance is attached to introducing in thermodynamic equations the most significant for natural science in whole concepts of flows $\mathbf{J}_i, \mathbf{j}_i$ as generalized rates of the transfer processes and the concept of power (capacity) of the energy conversion process in a whole system $N_i = \mathbf{X}_i \cdot \mathbf{J}_i$ and in its unit volume $\mathbf{x}_i \cdot \mathbf{j}_i$. It should be noted that the notion of capacity refers to only the useful energy conversion processes and, therefore, could not appear in the depths of the theory of irreversible processes restricted to consideration of exclusively dissipative phenomena. On the contrary, all basic relationships of this theory will hereinafter be obtained as a consequence from energodynamics.

This is enough in principle to construct a unitary theory of real processes enabling investigation of any systems (simplex and complex, closed and open, homogeneous and heterogeneous, isolated and non-isolated, tending to and omitting equilibrium) not outstepping the strict applicability of its primary concepts.

## References

1. *Etkin V.A.* Thermokinetics (Synthesis of Heat Engineering Theoretical Grounds). Haifa, 2010, 334 p.
2. *Clausius R.* Die mechanische Warmethorie. Braunschweig, 1876. Bd.1,2.
3. *Gelfer J.M.* History and Methodology of Thermodynamics and Statistical Physics. Edn 2.–M: Vysshaya shkola, 1981.
4. *Gyarmati I*. Non-Equilibrium Thermodynamics. Field Theory and Variation Principles. – Springer –Verlag, 1970.